\documentclass{article} 

\usepackage{lineno,hyperref}
\modulolinenumbers[5]

\usepackage{appendix}
\usepackage{amsmath,amssymb}
\usepackage{bm}
\usepackage{booktabs}
\usepackage{braket}
\usepackage{cite}
\usepackage{dcolumn}
\usepackage{graphicx}
\usepackage{hyperref}
\usepackage{multirow}
\usepackage[normalem]{ulem}

\numberwithin{equation}{section}

\begin{document}

\begin{center}
{\Large 
\textbf{A little hair can make a big difference:} \\ thermodynamic stability of quasi--bald asymptotically--flat black holes
}
\vspace{0.8cm}
\\
{Nuno M. Santos$^{\dagger, \star}$, Carlos A. R. Herdeiro$^{\dagger}$ and Eugen Radu$^{\dagger}$  
\\
\vspace{0.3cm}
$^{\dagger }${\small  Centre for Research and Development in Mathematics and Applications (CIDMA),}\\
{\small Department of Mathematics, University of Aveiro, 3810--193 Aveiro, Portugal}
\\
\vspace{0.3cm}
$^{\star}${\small Departamento de F\'\i sica,
Instituto Superior T\'ecnico -- IST, Universidade de Lisboa -- UL,} 
\\ {\small Avenida
Rovisco Pais 1, 1049--001 Lisboa, Portugal}
\vspace{0.3cm}
}
\end{center}
 
\begin{abstract}
The local thermodynamic stability of a black hole (BH) in the canonical ensemble is defined by the positivity of the specific heat at constant global charges. Schwarzschild BHs in thermodynamic equilibrium with an energy reservoir are always unstable against small fluctuations of energy, whereas sufficiently near--extremal Reissner-Nordstr\"{o}m/Kerr BHs  are stable. One could expect that asymptotically--flat hairy BHs branching off from such stable phases would also be, by continuity, locally thermodynamically stable for vanishingly little hair. We show this
is not the case in some models, including scalarized BHs bifurcating from Reissner--Nordstr\"{o}m and spinning BHs with
synchronized hair bifurcating from Kerr. Specifically, it is found that quasi--bald BHs are locally thermodynamically unstable in the canonical ensemble for all global charges and regardless of being dynamically and entropically preferred over bald ones at fixed global charges.
\end{abstract}




\section{\label{sec:1}Introduction}

In 1973, Bardeen, Carter and Hawking formulated the four laws of black--hole mechanics~\cite{Bardeen:1973gs} and noticed that the surface gravity $\kappa$ and area $A$ of the (spatial sections of the) event horizon of a stationary black hole (BH) bore a remarkable resemblance to temperature and entropy in classical thermodynamics, respectively. Soon after, Hawking made the remarkable discovery that BHs emit particles at a steady rate as if they were black bodies with temperature $T=\kappa/(2\pi)$ ~\cite{Hawking:1974sw,Hawking:1974rv}. 
These four laws of black--hole mechanics are not mere analogies with the standard laws of thermodynamics; they actually describe BHs as thermodynamic systems. It is therefore natural to ask whether BHs are thermodynamically stable or not.

 Thermodynamic stability can be local or global. \textit{Local stability} refers to whether a certain equilibrium phase of a system corresponds to a \textit{local} maximum of the entropy. It concerns the system's response to small fluctuations, determined by its thermodynamic variables under some fixed quantities, which amounts to a choice of ensemble. A system is said to be in a locally stable phase if any fluctuations produce a counteracting effect that ends up restoring the thermodynamic equilibrium. \textit{Global stability}, on the other hand, refers to whether a certain equilibrium phase of a system corresponds to a \textit{global} maximum of the entropy.

The local stability can be monitored by linear response functions such as the specific heat, $C$. The specific heat dictates how much a system's temperature changes when it absorbs heat from the environment. Consider, for instance, a Schwarzschild BH with mass $M$ at temperature $T=1/(8\pi M)$ in contact with a heat reservoir $R$ at fixed temperature $T_R$. Its specific heat is negative, $C=-1/(8\pi T)$. If $T<T_R$ (say), the BH will absorb energy from $R$. As a result, its temperature will decrease. Thus, the system runs away from thermal equilibrium and is (locally) unstable from a thermodynamic viewpoint. 

\begin{figure}[ht]
\centering
\includegraphics[width=0.5\columnwidth]{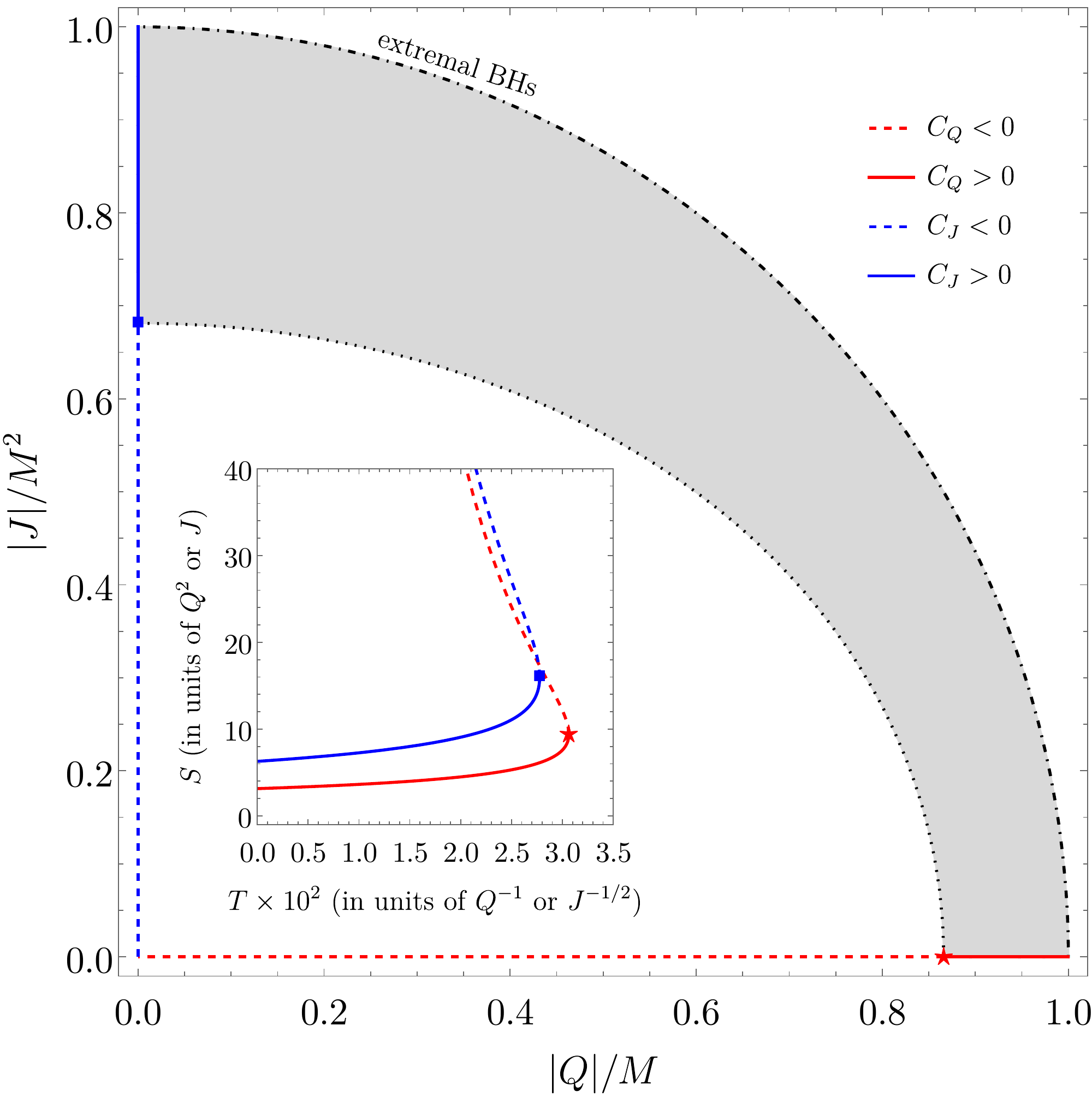}
\caption{
Local thermodynamic stability of Kerr--Newman BHs in the canonical ensemble. (Main panel) Sign of the specific heat at constant electric charge $Q$ and angular momentum $J$ in the plane $(|Q|/M,|J|/M^2)$: $C_{Q,J}<0$ in the (inner)  white region, whereas $C_{Q,J}>0$ in the gray region. (Inset) Entropy of Reissner--Nordstr\"{o}m and Kerr BHs as a function of their temperature. The markers in both plots refer to Reissner--Nordstr\"{o}m (red star) and Kerr (blue square) BHs with diverging specific heat.}
\label{fig:1}
\end{figure}

In general, however, the BH may have non--vanishing electric charge $Q$ and/or angular momentum $J$. Suppose now that a Kerr--Newman BH can exchange energy (at fixed temperature), but not electric charge nor angular momentum, with the reservoir (i.e. $Q$ and $J$ are kept fixed). This is the \textit{canonical ensemble}. The local thermodynamic stability is then characterized by the positivity of the specific heat at constant $Q$ and $J$,
\begin{align}
C_{Q,J}=\left(\frac{\partial M}{\partial T}\right)_{Q,J}=T\left(\frac{\partial S}{\partial T}\right)_{Q,J}\ ,
\end{align}
where $S$ is the BH entropy. As it turns out, there is a continuity with the Schwarzschild phase: the specific heat (at constant electric charge and angular momentum) is negative for sufficiently small $Q$ \textit{and} $J$. However, if \cite{Davies:1977bgr}
\begin{align}
J^4+6J^2M^4+4Q^2M^6-3M^8>0\ ,
\end{align}
it becomes positive and the system becomes thermodynamically locally stable (in this ensemble). This corresponds to the gray region in \autoref{fig:1}. For Reissner--Nordstr\"{o}m BHs ($J=0$), this occurs when $\sqrt{3}M/2<|Q|<M$ (horizontal red solid line). For Kerr BHs ($Q=0$), on the other hand, it holds for $\sqrt{2\sqrt{3}-3}M^2<|J|<M^2$ (vertical blue solid line). For ease of notation, hereafter $C_Q\equiv C_{Q,J=0}$ and $C_J\equiv C_{Q=0,J}$. The sign of the specific heat can be inferred from the curve $S=S(T)$ for fixed $Q$ and/or $J$. The inset of \autoref{fig:1} shows the entropy of Reissner--Nordstr\"{o}m and Kerr BHs as a function of their temperature. The black dotted line (in the main panel), together with the markers (in both panels), correspond to BHs with diverging specific heat, separating the stable and unstable phases. Such infinite discontinuity is commonly associated with second--order phase transitions.

\medskip

The above description exhausts the discussion concerning electrovacuum BHs in the canonical ensemble. In the last few years, however, a number of non--Kerr--Newman (but still asymptotically--flat) BHs became popular. They possess new macroscopic degrees of freedom not associated with gauge charges and collectively referred to as ``hair" -- see $e.g.$~\cite{Herdeiro:2015waa}. In particular, some of these models branch off from the Kerr--Newman family, thus being continuously connected to the electrovacuum BHs. In this context, one might wonder if a similar thermodynamic picture holds for asymptotically--flat hairy BHs continuously connected to BHs in general relativity (GR) -- i.e. \textit{are hairy BHs branching off from locally thermodynamically stable GR BHs also locally thermodynamically stable (in the same statistical ensemble)?}  By continuity, it seems intuitive that the answer should be positive for BHs with little hair. Indeed, this is what one observes in the Kerr--Newman family: adding $Q$ ($J$) to Kerr (Reissner-Nordstr\"om) BHs, the Kerr--Newman solutions retain a positive specific heat in the neighbourhood of Kerr (Reissner-Nordstr\"om) BHs with positive specific heat.  However, the addition of ``hair" (rather than global charges associated with gauge symmetries) can spoil the local thermodynamic stability of such electrovacuum  BHs. In other words, locally thermodynamically stable GR BHs may become unstable when they grow (even very little) hair. 

\medskip

Asymptotically--flat hairy BHs (in and beyond GR) may emerge from the reconsideration of the assumptions of no--hair theorems~\cite{Herdeiro:2015waa}.  In the theories of relevance here, bald and hairy BHs coexist and, at some scales, the former become unstable to forming hair and evolve into the latter. In other words, such hairy BHs have a dynamical formation mechanism and are continuously connected to GR BHs in the linear limit of the theories (when the hair is vanishingly little)~\cite{Herdeiro:2022yle}. To address the question of local thermodynamic stability of such hairy BHs, two illustrative families will be considered here, both defined by the action
\begin{align}
\mathcal{S}=\frac{1}{16\pi}\int\text{d}^4x\sqrt{-g}R+\mathcal{S}_\text{M}\ ,
\label{eq:1}
\end{align}
where $R$ is the Ricci scalar of the metric $g_{ab}$ with determinant $g$,  $\mathcal{S}_\text{M}=\int\text{d}^4x\mathcal{L}_\text{M}$ is the action for the matter field(s) and $\mathcal{L}_\text{M}=\sqrt{-g}\widehat{\mathcal{L}}_\text{M}$ is the corresponding Lagrangian density ($\widehat{\mathcal{L}}_\text{M}$ is a scalar). 

The first family comprises scalarized BHs in Einstein--Maxwell--scalar (EMs) theories~\cite{Herdeiro:2018wub,Astefanesei:2019pfq,Fernandes:2019rez,Fernandes:2019kmh,Blazquez-Salcedo:2020nhs}. These describe a massless real scalar field $\phi$ minimally coupled to Einstein's gravity and non--minimally coupled to Maxwell's electromagnetism, $\widehat{\mathcal{L}}_\text{M}=-2g^{ab}\phi_{;a}\phi_{;b}-f(\phi)\mathcal{I}$, where $\mathcal{I}=F_{ab}F^{ab}$, $F=\text{d}A$ is the Maxwell tensor and $f(\phi)$ is a coupling function. For a judicious choice of $f(\phi)$, EMs theories admit both GR and scalarized BHs. In particular, the former can undergo spontaneous scalarization and become hairy (similarly to neutron stars in scalar--tensor theories~\cite{Damour:1993hw}). 

The second family is composed of BHs with synchronized hair~\cite{Herdeiro:2014goa,Herdeiro:2016tmi}. They are found in theories featuring a massive complex bosonic field minimally coupled to Einstein's gravity: $\widehat{\mathcal{L}}_\text{M}=-\Psi^*_{;a}\Psi^{;a}-\mu^2|\Psi|^2$ for a scalar field $\Psi$ with mass $\mu$, whereas $\widehat{\mathcal{L}}_\text{M}=-F^{ab}F_{ab}^*/4-\mu^2A^aA_a^*/2$ for a vector field $A$ with mass $\mu$. Here, the asterisk denotes complex conjugation. In either case, BHs with synchronized hair coexist with Kerr BHs. At some scales, Kerr BHs become unstable against superradiance, which results in the transfer of energy and angular momentum to a bosonic cloud orbiting the BH~\cite{Herdeiro:2022yle}.  

It is convenient to sketch the similarities and differences between the two families of hairy BHs. Scalarized BHs bear some resemblance with BHs with synchronized hair in the sense that they emerge from the growth and saturation of an instability. The instability (tachyonic for scalarized BHs and superradiant for BHs with synchronized hair) is present in the linear limit of the theory and is responsible for the development of hair when non--linear effects are taken into account. Besides, both types of BHs are dynamically preferred over their bald counterparts. In fact, they are entropically favoured, i.e. they maximize the entropy of the system in the microcanonical ensemble, $i.e.$ for fixed global charges $(M,Q,J)$.

There are, however, some differences one should remark. While BHs with synchronized hair reduce to bosonic stars in the limit of vanishing horizon size, some EMs theories of scalarized BHs do not possess solitons, as established by some no--go theorems~\cite{Herdeiro:2019oqp}. Even when they do, the solitons may not be continuously connected with the hairy BHs~\cite{Dias:2021vve}. Another important distinction between the two families concerns the symmetries of the bosonic field. In the EMs theories, the scalar field shares the symmetries with the spacetime. As for BHs with synchronized hair, although the spacetime is stationary and axi--symmetric, the bosonic field depends explicitly on time (but, since it is complex, the corresponding energy--momentum tensor is time--independent). Finally, BHs with synchronized hair appear in models with a global $U(1)$ symmetry, which makes the hair \textit{primary}, measured by a conserved (in the sense of a continuity equation) Noether charge. By contrast, the hair in scalarized BHs is \textit{secondary} and the corresponding scalar ``charge" is not conserved in any meaningful sense.

The thermodynamics of asymptotically--flat hairy BHs is still poorly explored. Most studies focus on the thermodynamic stability in the microcanonical ensemble. The purpose of this paper is to provide an investigation of the local thermodynamic stability of the aforementioned familes of hairy BHs in the canonical ensemble. Their specific heat (at constant global charges) is computed numerically and found to be negative for quasi--GR BHs, regardless of their specific electric charge or angular momentum.


\section{\label{sec:2}Scalarized black holes}

In EMs theories, the equation of motion for the scalar field reads $\Box\phi=f_{,\phi}\mathcal{I}/4$. $\phi=0$ solves the equation of motion if $f_{,\phi}(0)=0$, in which case GR BHs remain solutions. However, they are not unique in general and coexist with BHs with a non--trivial scalar field (or \textit{scalar hair}). These are usually dubbed \textit{scalarized BHs}. 

One requires scalarized BHs to be continuously connected to GR BHs (i.e. the former reduce to the latter in the linear limit of the theory). These fall into Subclass IIA in~\cite{Astefanesei:2019pfq}. Such bifurcation may arise when GR BHs are afflicted by a linear tachyonic instability. The linearized Klein--Gordon equation reads $(\Box-\mu_\text{eff}^2)\phi=0$, with $\mu_\text{eff}^2=f_{,\phi\phi}(0)\mathcal{I}/4$. The coupling function together with the source term $\mathcal{I}$ act like a negative contribution to the field's mass provided that $f_{,\phi\phi}(0)\mathcal{I}<0$. In that case, GR BHs become unstable to growing hair. For a purely electric field, $\mathcal{I}<0$ and the previous inequality reduces to $f,_{\phi\phi}(0)>0$. Some possible choices for the coupling function are then \cite{Astefanesei:2019pfq}: \textit{exponential coupling}, $f_E(\phi)=e^{-\alpha\phi^2}$ \cite{Herdeiro:2018wub,Fernandes:2019rez}; \textit{hyperbolic cossine coupling}, $f_C(\phi)=\cosh(\sqrt{-2\alpha}\phi)$ \cite{Fernandes:2019rez}; \textit{power coupling}, $f_P(\phi)=1-\alpha\phi^2$ \cite{Boskovic:2018lkj,Fernandes:2019rez}. $\alpha$ will be referred to as the coupling constant and must be negative so that $f_{,\phi\phi}(0)>0$. When $\alpha<0$, $f_E$, $f_C$ and $f_P$ are monotonically increasing functions of $\phi$.

Any static, spherically--symmetric solution to the equations of motion can be cast in the form
\begin{align}
\text{d}s^2=-N(r)e^{-2\delta(r)}\text{d}t^2+\frac{\text{d}r^2}{N(r)}+r^2(\text{d}\theta^2+\sin^2\theta\text{d}\varphi^2)\ ,
\end{align}
in Schwarzschild coordinates $(t,r,\theta,\varphi)$, where $N(r)\equiv 1-2m(r)/r$ and $m(r)$ is the Misner--Sharp mass function, which can be regarded as the quasi--local mass contained within a sphere of radius $r$. Spherical symmetry imposes an electrostatic $4$--vector potential (in the absence of a magnetic charge) as well as a radial--dependent scalar field, i.e. $A_a\text{d}x^a=V(r)\text{d}t$ and $\phi=\phi(r)$, where $V$ is the electrostatic potential.

One assumes the existence of an event horizon at 
$r=r_H$, which is the largest root of $N$. The boundary conditions for $m$, $\delta$, $V$ and $\phi$ at the event horizon are found by requiring them to have a regular Taylor series at $r=r_H$, with $r_H=2m(r_H)$, $\delta(r_H)=\delta_0$, $V(r_H)=0$ and $\phi(r_H)=\phi_0$, where the gauge condition $V(r_H)=0$ was imposed. In addition to $\{\alpha,\delta_0,\phi_0\}$, the solutions are characterized by the Arnowitt--Deser--Misner (ADM) mass $M$, the electric charge $Q$, the scalar charge $Q_s$ and the (asymptotic) electrostatic potential $\Phi$. When $\phi=0$, all non--singular (on and outside the event horizon) asymptotically--flat, electrically charged BHs belong to the Reissner--Nordstr\"{o}m family of BHs. 

The BH entropy and temperature are
\begin{align}
S=\pi r_H^2 \ ,
\quad
T=\frac{e^{-\delta_0}}{4\pi r_H}\left[1-\frac{Q^2}{r_H^2f(\phi_0)}\right]\ .
\end{align}
Since the temperature cannot be negative, $r_H^2f(\phi_0)\geqslant Q^2$. The mass $M$, the electric charge $Q$, the electrostatic potential $\Phi$, the entropy $S$ and the temperature $T$ satisfy the Smarr relation $M=2TS+\Phi Q$, the first law of BH mechanics being $
\text{d}M=T\text{d}S+\Phi\text{d}Q$, and, what's more, the non--linear relation $M^2+Q_s^2=4T^2S^2+Q^2$.

An important feature of these scalarized BHs is that the electric charge $Q$ does not necessarily coincide with the electric charge \textit{on} the event horizon  $Q_H$. The latter is simply $Q_H=Q/f(\phi_0)$, which suggests that the non--minimal coupling between the scalar field and the 4--vector porential results in electric charge outside the event horizon. It is convenient to introduce the ``hairiness" parameter
\begin{align}
h\equiv1-\frac{Q_H}{Q}=1-\frac{1}{f(\phi_0)}\ ,
\label{eq:2.3}
\end{align}
which measures (through $\phi_0$) the fraction of electric charge outside the event horizon (or, loosely speaking, how hairy a scalarized BH is). Note that $h\in[0,1)$. In the linear limit of the theory ($\phi_0=0$), $f(\phi_0)=1$, $Q=Q_H$ and $h=0$.

\subsection{Tachyonic instability and domain of existence} 
When $f_{,\phi\phi}(0)>0$, the source term $\mathcal{I}=F_{ab}F^{ab}$ provides the scalar field with an imaginary (effective) mass, triggering a tachyonic instability. Such instability may arise even for a test field $\phi=\delta\phi$, with $|\delta\phi|M\ll 1$, turning Reissner--Nordstr\"{o}m BHs unstable. 

The scalar--free solutions at the onset of the tachyonic instability form an \textit{existence line} on the plane defined by $(\alpha,|Q|/M)$. Since the second--order Taylor expansion of $f_E$ and $f_C$ around $\phi=0$ coincides with $f_P$, the existence line is the same for the coupling functions considered herein. For each $\alpha$, scalarized BHs branch off from the existence line and always have a charge--to--mass $|Q|/M$ ratio greater than that of the Reissner--Nordstr\"{o}m BH at the bifurcation point. Moreover, they can exceed the usual extremal limit and have $Q^2>M^2$. The solution space of scalarized BHs for the different couplings is qualitatively similar and can be found in~\cite{Fernandes:2019rez}. Besides the existence line, it is bounded by a \textit{critical line} comprised of singular solutions, with vanishing entropy.

\subsection{\label{sec:2.3}Local thermodynamic stability}
As  mentioned in \autoref{sec:1}, Reissner--Nordstr\"{o}m BHs with sufficiently small (large) electric charge--to--mass ratio $Q/M$ are locally thermodynamically unstable (stable) in the canonical ensemble, $i.e.$ their specific heat at constant electric charge $C_Q$ is negative (positive). The state of affairs changes with the addition of a scalar field non--minimally--coupled to electromagnetism to the theory. Scalarized BHs always have negative specific heat (at constant electric charge) in the linear limit, i.e. when $\phi_0\rightarrow0$, as we shall now discuss.
This means scalarized BHs cannot be a local minimum of the action and therefore must have a negative mode~\cite{Reall:2001ag}, despite being dynamically preferred. 

\begin{figure}[ht]
\centering
\includegraphics[width=0.475\columnwidth]{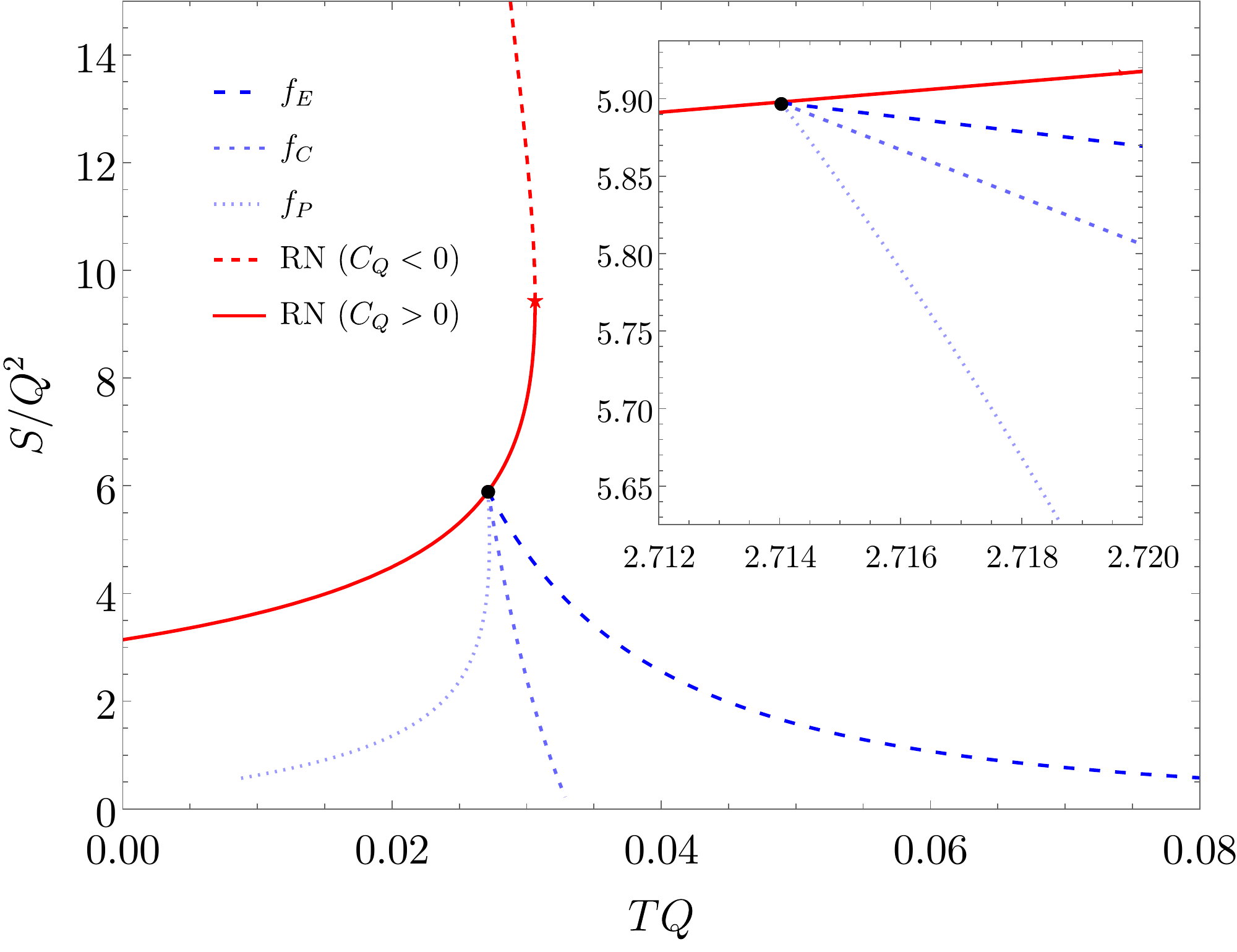}~
\includegraphics[width=0.475\columnwidth]{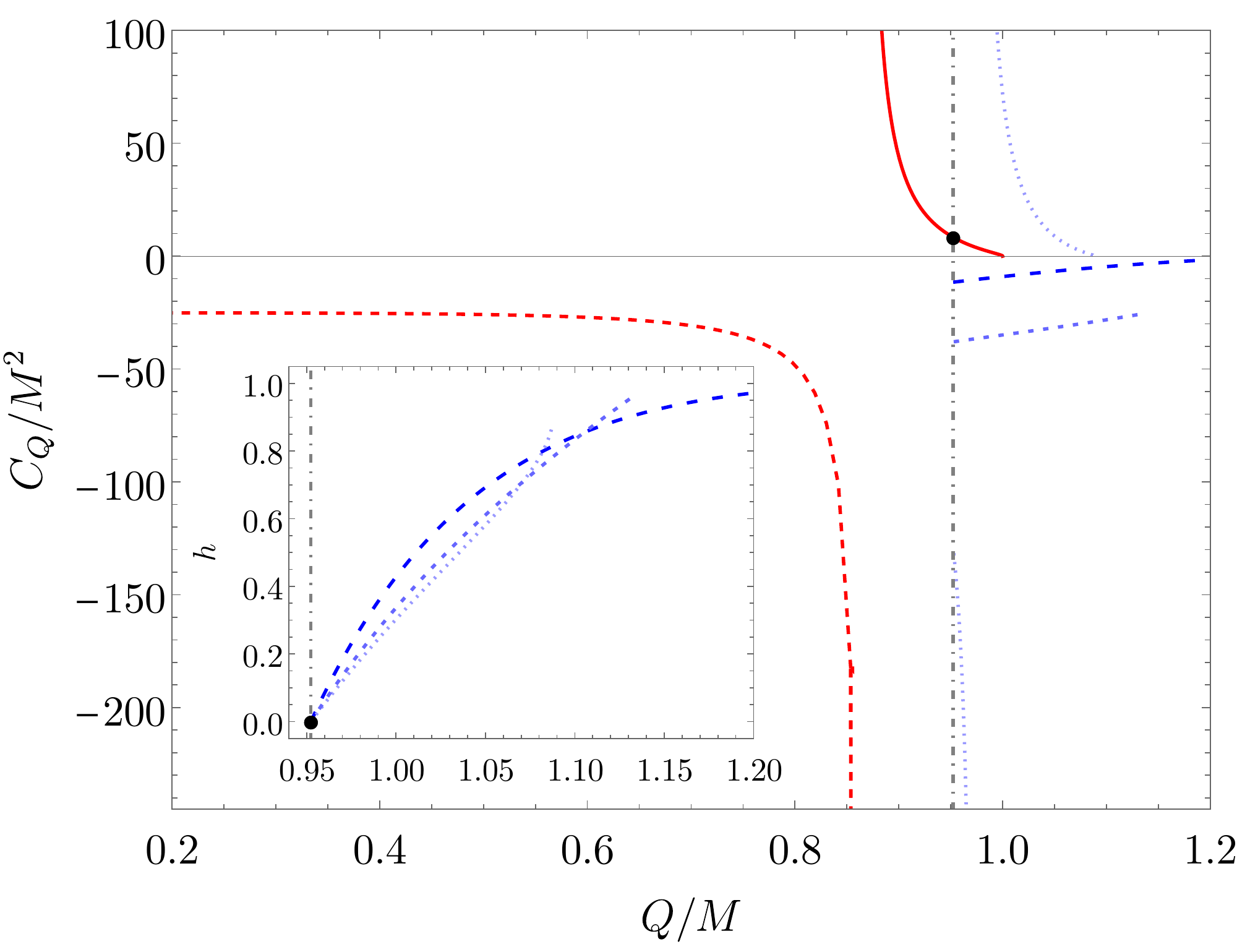}
\caption{(Left panel) Entropy of the scalarized BHs with $Q=0.4$ as a function of their temperature, for different coupling functions. The inset shows the behavior of the curves close to the bifurcation point (black dot). (Right panel) Specific heat at constant electric charge of Reissner--Nordstr\"{o}m BHs with $Q=0.4$ as a function of their charge--to--mass ratio $Q/M$. The vertical dotdashed line corresponds to the charge--to--mass ratio of the bifurcation point. The inset shows the hairiness $h$ defined in Eq.~\eqref{eq:2.3}}.
\label{fig:2}
\end{figure}

The blue lines in both panels of \autoref{fig:2} correspond to scalarized BHs with $Q=0.4$ for the different coupling functions. The value of the coupling constant ($\alpha=-2$) was chosen so that scalarized BHs bifurcate from locally thermodynamically stable Reissner--Nordstr\"{o}m BHs. $\phi_0$ (and thus $h$) increases monotonically as one moves down along the blue lines (see inset in the right panel of~\autoref{fig:2}), signalling hairier and hairier BHs. The lines appear to terminate at singular solutions with vanishing entropy. The limiting behavior of the temperature differ: it appears to diverge for $f=f_E$, to tend to a non--vanishing finite value for $f=f_C$ and to vanish for $f=f_P$.\footnote{In the latter case, because the temperature starts decreasing as the scalarized BHs become smaller and smaller, one could say that their behavior is akin to that of extremal Reissner--Nordstr\"{o}m  BHs. However, it should be kept in mind that extremal Reissner--Nordstr\"{o}m BHs, having a finite entropy, are not singular. One can show in particular that these theories do not admit the near--horizon geometry of extremal Reissner--Nordstr\"{o}m BHs as a solution.} This distinction is connected to the higher--order terms in $\phi^2$ of the Taylor expansion of $f_E$ and $f_C$.  

At first sight, the left panel of~\autoref{fig:2} may suggest that the specific heat is negative for scalarized BHs with exponential and hyperbolic cossine coupling to electromagnetism, but positive for those with a power coupling. On closer inspection (see inset in the left panel of~\autoref{fig:2}), though, it becomes clear that the latter has negative specific heat in the linear limit. This indicates a second--order phase transition. The specific heat at constant electric charge $Q=0.4$ of Reissner--Nordstr\"{om} and scalarized BHs is shown in the right panel of~\autoref{fig:2}. As the charge--to--mass ratio approaches its maximum value, it vanishes for $f=f_E,f_P$ and tends to a non--vanishing finite value for $f=f_C$.   

Consider the case of the Reissner--Nordstr\"{o}m BH at the point of intersection of the two sets of solutions (black dot in~\autoref{fig:2}). This BH has $C_Q>0$. Suppose that it absorbs a small amount of positive energy $\delta M>0$ (without exchanging electric charge). The total mass $M$ of the BH will increase by $\delta M$, thus reducing $|Q|/M$. The BH moves up along the red solid line, preserving its local thermodynamic stability. The event horizon absorbs all the energy so that the scalar field remains trivial. If, on the contrary, the Reissner--Nordstr\"{o}m BH absorbs a small amount of negative energy $\delta M<0$, $M$ will decrease, yielding a BH with higher $|Q|/M$. This could either be a Reissner--Nordstr\"{o}m BH or a scalarized BH. In general, in the region where the Reissner--Nordstr\"{o}m and scalarized BHs co--exist, the scalarized solutions maximize the entropy and thus are thermodynamically favoured. The negative energy feeds the field, triggering the tachyonic instability, which results in a non--trivial field in equilibrium with the BH. Further absorption of negative energy enhances the tachyonic instability so that the BH migrates downards along the blue curves in \autoref{fig:2} (left panel). 


\section{\label{sec:3}Black holes with synchronized hair}

BHs with synchronized hair are four--dimensional, asymptotically--flat, stationary solutions of Einstein’s gravity minimally coupled to a complex bosonic field $\psi$ with non--vanishing mass $\mu$. As solutions describing BHs, they feature an event horizon at $r=r_H$, being regular on and outside it. The matter field is characterized by a harmonic time and azimuthal dependence, $\psi\sim e^{-i\omega t+im\varphi}$, where $\omega>0$ and $m\in\mathbb{Z}^+$ are its frequency and azimuthal harmonic index, respectively. The ansatz and corresponding equations of motion as well as the boundary condition at the event horizon, spatial infinity and on the axis can be found in~\cite{Herdeiro:2015gia} for scalar hair ($\psi=\Phi$) and in~\cite{Herdeiro:2016tmi} for vector hair ($\psi=A$). 

\subsection{\label{sec:3.1}Superradiant instability and domain of existence}

This family of BHs is continuously connected to (a subset of) the family of Kerr BHs. This means BHs with synchronized hair can be realized in the linear limit of the theory, in which the backreaction of the spacetime to a non--constant bosonic field is negligible. The corresponding limiting solutions are bound states between Kerr BHs and non--trivial bosonic fields, commonly known as \textit{stationary clouds}. 

Stationary clouds are nothing but zero modes of the \textit{superradiant instability}, i.e. equilibrium states defined by a bosonic field with vanishing momentum near the event horizon, which amounts to the \textit{synchronization condition}
\begin{align}
\Omega_H=\frac{\omega}{m}\ ,
\label{eq:3.1}
\end{align}
where $\Omega_H$ denotes the BH angular velocity. Eq. \eqref{eq:3.1} sets the onset  of superradiance in Kerr BHs, which occurs whenever $\omega<m\Omega_H$. This condition follows directly from the first and second laws of BH mechanics. BH superradiance is rooted in the existence of an ergoregion, within which negative--energy physical states are possible. However, such possibility does not automatically translate into an instability, unless a confinement mechanism is present. This is here naturally provided by the bare mass of the bosonic field. 

Linearizing the equations of motion around $\psi=\psi_0$, for some constant $\psi_0$, one can show that, when $\omega/\Omega_H$ is not an integer, the field perturbation $\delta\psi\equiv(\psi-\psi_0)$ has a non--vanishing momentum and oscillates in space, namely in the radial direction, near the event horizon. On the contrary, when $\omega/\Omega_H$ is an integer, the field becomes stationary and binds to the Kerr BH to form a stationary cloud, pretty much like an electron in a hydrogen atom.

When the backreaction of the spacetime is taken into account and Eq.~\eqref{eq:3.1} is satisfied, (some) Kerr BHs grow hair and turn dynamically into BHs with synchronized hair. These solutions were originally found for free scalar \cite{Herdeiro:2014goa} and vector \cite{Herdeiro:2016tmi} fields, and later generalized for self--interacting fields and/or non--minimal couplings \cite{Herdeiro:2015tia}. 

\begin{figure}[ht]
\centering
\includegraphics[width=0.475\columnwidth]{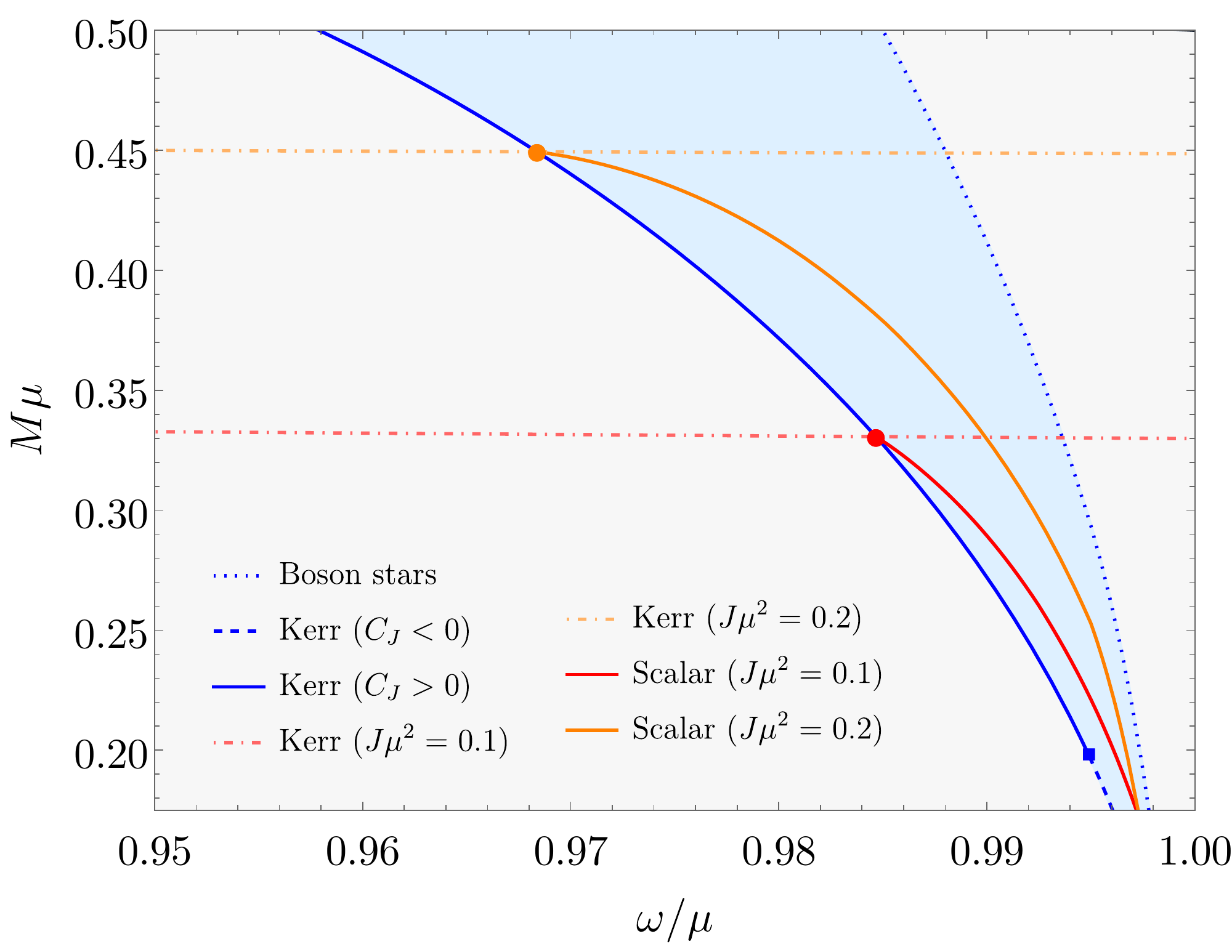}~
\includegraphics[width=0.475\columnwidth]{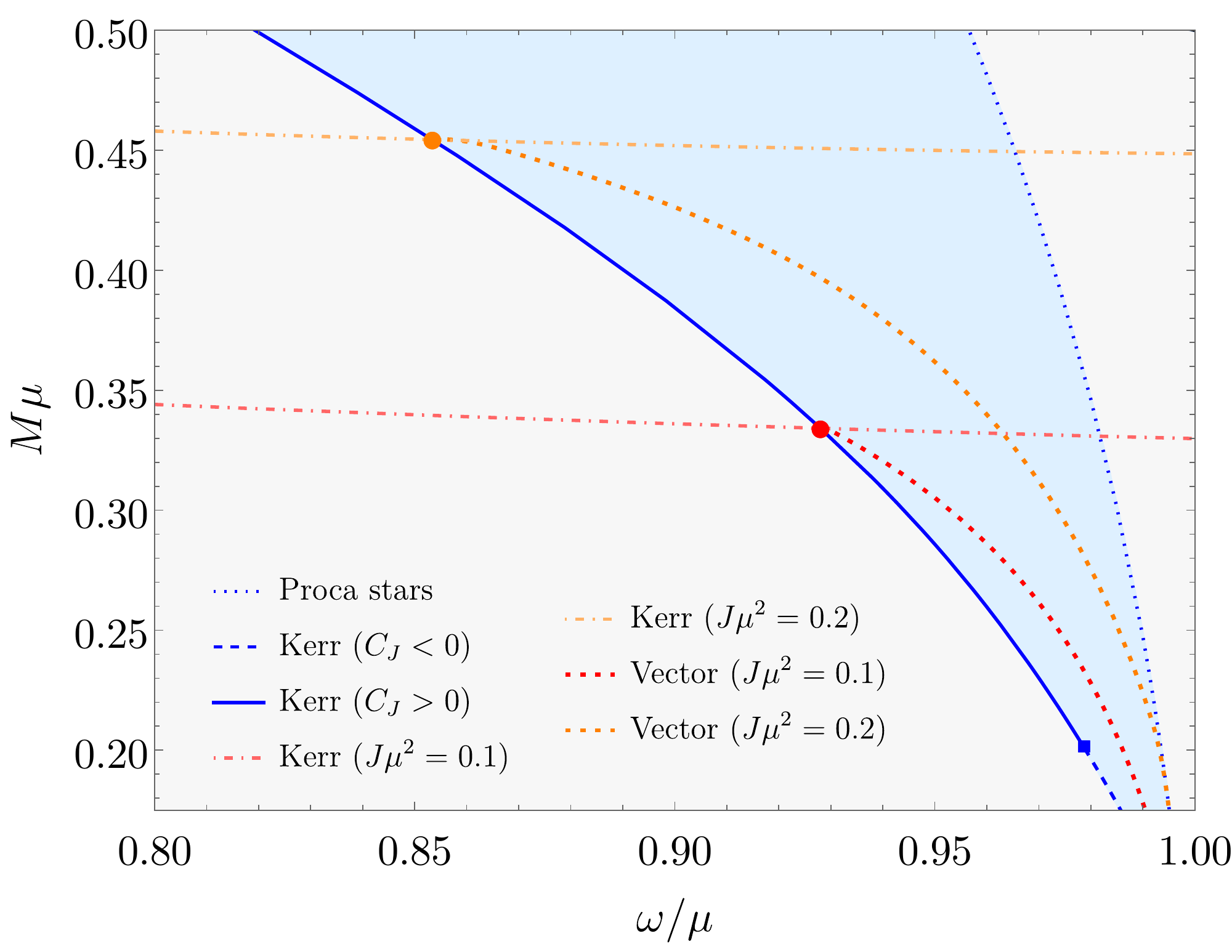}
\caption{Domain of existence of BHs with synchronized scalar (left
panel) and vector (right panel) hair with $(n,m)=(0,1)$ in the $(\omega/\mu,M\mu)$ plane (light blue region). Kerr BHs live in the light gray region. The blue square separates Kerr BHs with $|J|<\sqrt{2\sqrt{3}-3}M^2$ ($C_J<0$, dashed blue line) and $|J|>\sqrt{2\sqrt{3}-3}M^2$ ($C_J>0$, solid blue line). The red and orange solid lines define BHs with synchronized hair with $J\mu^2=0.1$ and $J\mu^2=0.2$, respectively. The circles represent the bifurcation points (cf. \autoref{tab:1}).}
\label{fig:3.1}
\end{figure}

The solution space of BHs with synchronized hair is fully described by $(i)$ two continuous dimensionless parameters, namely the ADM mass, $M\mu$, and the oscillation frequency, $\omega/\mu$, or, equivalently, the ADM angular momentum $J\mu^2$, in units of the field's mass; $(ii)$ two discrete parameters, namely the number of nodes in the radial direction, $n\in\mathbb{N}_0$, and the azimuthal harmonic index, $m\in\mathbb{Z}^+$. The solutions live in a subset of the plane defined by $(\omega/\mu,M\mu)$ (say). Fixing $(n,m)$, they populate a spiral--shaped region, regardless of the spin of the bosonic field, a part of which is shown in \autoref{fig:3.1} for BHs with synchronized scalar (left panel) and vector (right panel) hair with $(n,m)=(0,1)$. They belong to the fundamental family of solutions, characterized by the lowest maximum ADM mass. The solution space is bounded by: $(i)$ the \textit{existence line}, comprised of stationary clouds around Kerr BHs (solutions with vanishing field); $(ii)$ the \textit{solitonic line}, comprised of spinning bosonic stars (solutions with vanishing horizon). BHs with synchronized hair interpolate between these two families of limiting solutions. 

The global charges $M$ and $J$, defined by Komar integrals, can be expressed as $M=M_H+M_{\psi}$ and $J=J_H+J_\psi$, where $M_H$ and $J_H$ ($M_\psi$ and $J_\psi$) are the mass and angular momentum inside (outside) the event horizon, respectively. As before, it is convenient to have some measure of the hairiness of these solutions. These can be the proportion of energy and angular momentum in the bosonic field
\begin{align}
p\equiv \frac{M_\psi}{M}\ ,
\quad
q\equiv \frac{J_\psi}{J}\ ,
\end{align}
respectively, where $p,q\in[0,1]$. Stationary clouds (bosonic stars) are characterized by $p=q=0$ ($p=q=1$).

\subsection{\label{sec:3.2}Local thermodynamic stability}
Just like Reissner--Nordstr\"{o}m BHs, Kerr BHs with sufficiently large (small) specific angular momentum $J/M^2$ are locally thermodynamically stable (unstable) in the canonical ensemble. Given the similarities and differences between scalarized BHs and BHs with synchronized hair, \textit{a priori}, it is not clear if the latter are also locally thermodynamically unstable in the canonical ensemble. Here, local thermodynamic stability is equivalent to the positivity of the specific heat at constant angular momentum $C_J$.

\begin{figure}[ht]
\centering
\includegraphics[width=0.475\columnwidth]{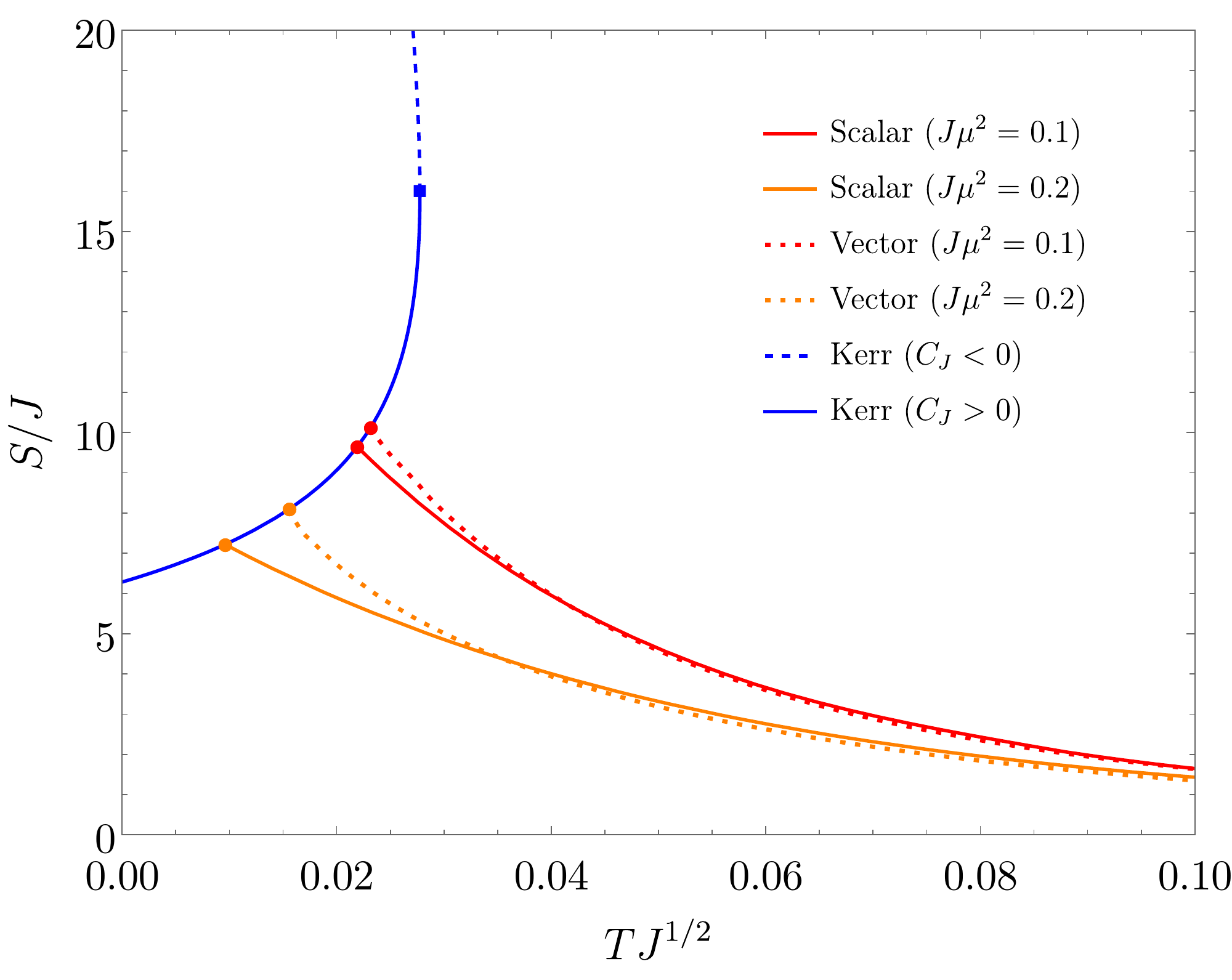}~
\includegraphics[width=0.475\columnwidth]{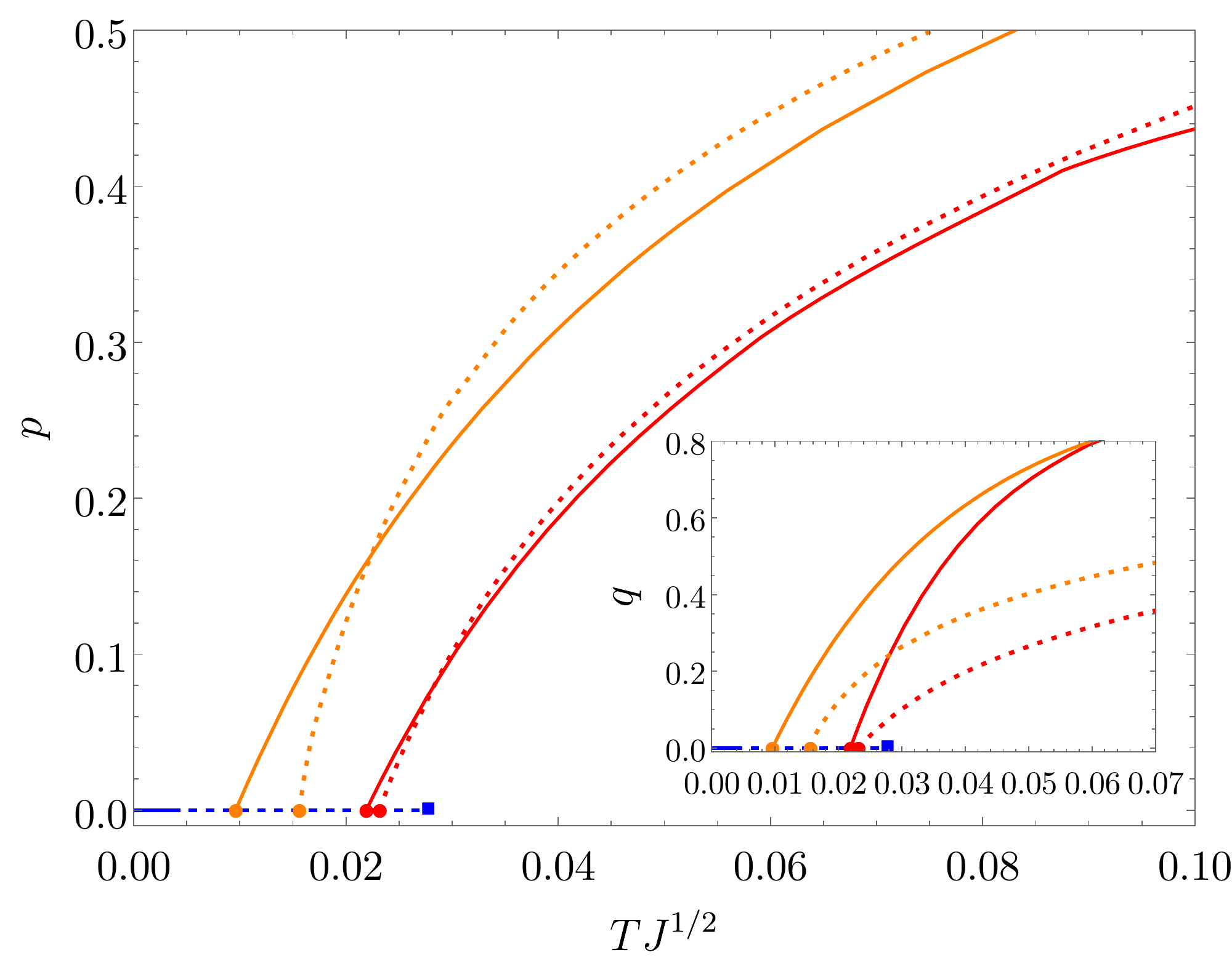}
\caption{Entropy (left panel) and hairiness $p$ and $q$ (right panel) of BHs with synchronized scalar and vector hair with $(n,m)=(0,1)$ and $J\mu^2=0.1$ (red lines) and $J\mu^2=0.2$ (orange lines) as a function of their temperature.}
\label{fig:3.2}
\end{figure}

BHs with synchronized hair of constant angular momentum $J\mu^2$ define a line segment in the domain of existence joining the Minkowski limit $(M,J\rightarrow0)$ to the existence line $(p,q\rightarrow0)$. As $J\mu^2$ increases, the bifurcation point approaches the extremal line ($p=q=0$, $J/M^2=1$). \autoref{fig:3.1} shows BHs with synchronized scalar (solid lines) and vector (dotted lines) hair with $J\mu^2=0.1$ (red lines) and $J\mu^2=0.2$ (orange lines). They bifurcate from locally thermodynamically stable Kerr BHs (blue solid line), since $|J|>\sqrt{2\sqrt{3}-3}M^2\approx0.6813M^2$ (see \autoref{tab:1}). However, as in~\autoref{sec:2.3}, the hairy BHs turn out to be unstable. Indeed, the left panel of~\autoref{fig:3.2} shows their entropy decreases as the temperature increases, which means that $C_J<0$. The hairiness $p$ increases as the temperature increases (see right panel of~\autoref{fig:3.2}). In the Minkowski limit, $S$ vanishes and $T$ diverges, with $C_J$ approaching zero. This behavior bears close resemblance to that of scalarized BHs in EMs theories with an exponential coupling (see~\autoref{fig:2}).
\begin{table}[h!]
\centering
\begin{tabular}{c||c||cccc}
& $J\mu^2$ & $M\mu$ & $\omega/\mu$ & $M\omega$ & $J/M^2$ \\ \midrule
\multirow{2}{*}{Scalar} & $0.10$ & 0.3307 & 0.9847 & 0.3256 & 0.9146\\
& $0.20$ & 0.4494 & 0.9684 & 0.4352 & 0.9904 \\
 \midrule
\multirow{2}{*}{Vector} & $0.10$ & 0.3341 & 0.9280 & 0.3101 & 0.8958\\
& $0.20$ & 0.4544 &  0.8534 & 0.3878 & 0.9686\\
\end{tabular}
\caption{Bifurcation points of BHs with synchronized hair with $(n,m)=(0,1)$ for the values of $J\mu^2$ presented in~\autoref{fig:3.1}.
}
\label{tab:1}
\end{table}

\section{\label{sec:4}Conclusion}
The issue of thermodynamical stability of BHs does not directly impact on their astrophysical viability. For instance, Schwarzschild and slowly rotating Kerr BHs are thermodynamically unstable in the canonical ensemble. Yet, they may exist in the Cosmos. 

In fact, considering the grand--canonical ensemble, where BHs are also allowed to exchange electric charge (at fixed electrostatic potential) or angular momentum (at fixed angular velocity), all electrovacuum BHs are locally thermodynamically unstable. This is because another response function becomes negative precisely in the region where the specific heat becomes positive for both the Reissner--Nordstr\"{o}m and Kerr cases~\cite{Davies:1977bgr}. 

Still, the issue of thermodynamic stability of BHs is relevant, and quite fruitful, for instance in the context of AdS/CFT~\cite{Witten:1998zw}. Thus, it becomes an interesting question to understand how adding extra properties, such as ``hair", to a BH affects its thermodynamic stability.

This paper addressed the local thermodynamic stability in the canonical ensemble of two families of BHs with bosonic hair, namely scalarized BHs in a subclass of EMs theories and BHs with syncrhonzied hair. Both these families are continuously connected to electrovacuum BHs and, therefore, provide examples where the BH hair can be arbitrarily small. Moreover, the former (latter) yields an example of secondary (primary) hair.

By studying their corresponding specific heat, it was found that the addition of a bosonic field minimally coupled to Einstein's gravity can change the thermodynamic behaviour of BHs, even when the field strength is vanishingly small. Specifically, quasi--bald BHs are locally unstable in this statistical ensemble, regardless of their specific global charges (electric charge or angular momentum). This is particularly surprising for hairy BHs branching off from thermodynamically stable GR BHs. 

This analysis provides a contrast between thermodynamical stability and dynamical stability. The Reissner-Nordstr\"om (Kerr) BHs with high specific electric charge (angular momentum) are the ones that are locally thermodynamically stable in the canonical ensemble. But when we enlarge the model to include the new fields and couplings of the models above, they are simultaneously the BHs more prone to the tachyonic (superradiant) instability that leads to the branching off towards the hairy solutions. The latter correspond to a new phase in which the BHs are thermodynamically locally unstable in the canonical ensemble, despite being entropically favoured for fixed global charges and therefore the ones preferred in a conservative dynamical evolution.

Although the analysis herein was restricted to the aforesaid families, similar results were found for vectorized Reissner--Nordstr\"{o}m BHs in Einstein--Maxwell-vector theories~\cite{Oliveira:2020dru} in a preliminary investigation, suggesting a \textit{universal behavior}. It would be interesting to perform a similar analysis on other families of hairy BHs continuously connected to the electrovacuum BHs of GR, e.g. scalarized Schwarzschild~\cite{Doneva:2017bvd,Silva:2017uqg} or Kerr BHs\cite{Cunha:2019dwb} in the extended scalar--tensor--Gauss--Bonnet theory. We emphasise that our two examples cover both primary and secondary hair, and that the observed behaviour contrasts with the one observed when adding global charges, $e.g.$ adding $Q$ to Kerr or $J$ to Reissner-Nordstr\"om, where a continuity in the thermodynamic stability properties is observed.

Besides considering the thermodynamic behaviour in the canonical ensemble, one can also examine the stability of the hairy BHs in the grand--canonical ensemble. In this statistical ensemble, the electrostatic potential $\Phi$ and the angular velocity $\Omega_H$ of the event horizon are fixed and the electric charge $Q$ and the angular momentum $J$ are free to vary. This is the most generic physical scenario. In this ensemble the whole Kerr--Newman family is unstable. The corresponding stability analysis of hairy BHs continuously connected to the Kerr--Newman family is left for future work.


\section*{Acknowledgements}
NS would like to thank Alexandre M. Pombo for sharing his knowledge on scalarized BHs and Jorge F. M. Delgado for fruitful discussion about BHs with synchronized hair.

This work is supported by the Center for Research and Development in Mathematics and Applications (CIDMA) through the Portuguese Foundation for Science and Technology (FCT -- Funda\c{c}\~ao para a Ci\^encia e a Tecnologia), references UIDB/04106/2020 and UIDP/04106/2020, and by national funds (OE), through FCT, I.P., in the scope of the framework contract foreseen in the numbers 4, 5 and 6 of the article 23, of the Decree-Law 57/2016, of August 29, changed by Law 57/2017, of July 19. The authors acknowledge support from the projects PTDC/FIS-OUT/28407/2017, CERN/FIS-PAR/0027/2019, PTDC/FIS-AST/3041/2020 and CERN/FIS-PAR/0024/2021. This work has further been supported by  the  European  Union's  Horizon  2020  research  and  innovation  (RISE) programme H2020-MSCA-RISE-2017 Grant No.~FunFiCO-777740.

\bibliographystyle{jhep}  
\bibliography{references}

\end{document}